\documentclass[5p]{elsarticle}

\usepackage{lineno,hyperref,url}
\modulolinenumbers[5]

\usepackage{amsmath}
\hypersetup{colorlinks,allcolors=black}
\journal{PRC}

\begin{document}

\begin{frontmatter}

\title{Direct determination of the excitation energy of quasi-stable isomer $^{180m}$Ta}


\author[a]{D.A.~Nesterenko}
\author[b]{K.~Blaum}
\author[c]{P.~Delahaye}
\author[b]{S.~Eliseev}
\author[a]{T.~Eronen}
\author[b]{P.~Filianin}
\author[d]{Z.~Ge}
\author[a]{M.~Hukkanen}
\author[a]{A.~Kankainen}
\author[e]{Yu.N.~Novikov}
\author[e]{A.V.~Popov}
\author[a]{A.~Raggio}
\author[a]{M.~Stryjczyk}
\author[a]{V.~Virtanen}

\address[a]{University of Jyv\"askyl\"a, Accelerator Laboratory, Department of Physics, University of Jyv\"askyl\"a, Finland}
\address[b]{Max-Planck Institut f. Kernphysik, Heidelberg, Germany}
\address[c]{GANIL, CEA/DSM-CNRS/IN2P3, Bd Henri Becquerel, 14000 Caen, France}
\address[d]{GSI Helmholtzzentrum f. Schwerionenforschung, Darmstadt, Germany}
\address[e]{NRC ``Kurchatov Institute'' - Petersburg Nuclear Physics Institute, Gatchina, Russia}

\begin{abstract}
$^{180m}$Ta is a naturally abundant quasi-stable nuclide and the longest-lived nuclear isomer known to date. It is of interest for, among others, the search for dark matter, for the development of a gamma laser and for astrophysics. So far, its excitation energy has not been measured directly but has been based on an evaluation of available nuclear reaction data. We have determined the excitation energy of this isomer with high accuracy using the Penning-trap mass spectrometer JYFLTRAP. The determined mass difference between the ground and isomeric states of $^{180}$Ta yields an excitation energy of 76.79(55) keV for $^{180m}$Ta. This is the first direct measurement of the excitation energy and provides a better accuracy than the previous evaluation value, 75.3(14) keV.

\end{abstract}

\begin{keyword}
\texttt Penning trap \sep high-precision mass spectrometry \sep nuclear isomers
\end{keyword}

\end{frontmatter}

\section{Introduction}

Nuclear isomers were discovered around 100 years ago \cite{Walker2020}. Today, around 1938 excited isomeric ($T_{1/2} \geq 100$ ns) states are known \cite{AME-20}. $^{180m}$Ta isomer with spin-parity $J^{\pi} = 9^-$ and excitation energy of around 75 keV is a unique nuclear isomer. The half-life of $^{180m}$Ta is longest of all isomeric states, longer than $4.5 \times 10^{16}$ y [2]. The isomer is much longer-living than the $J^{\pi} = 1^+$ ground state of $^{180}$Ta, which has a half-life of 8.1 h. As a consequence, $^{180}$Ta is present in nature in its isomeric state, making a 0.01201(8)\% fraction of natural tantalum. 

\begin{figure}[t]
\centering
\includegraphics[width=0.4\textwidth]{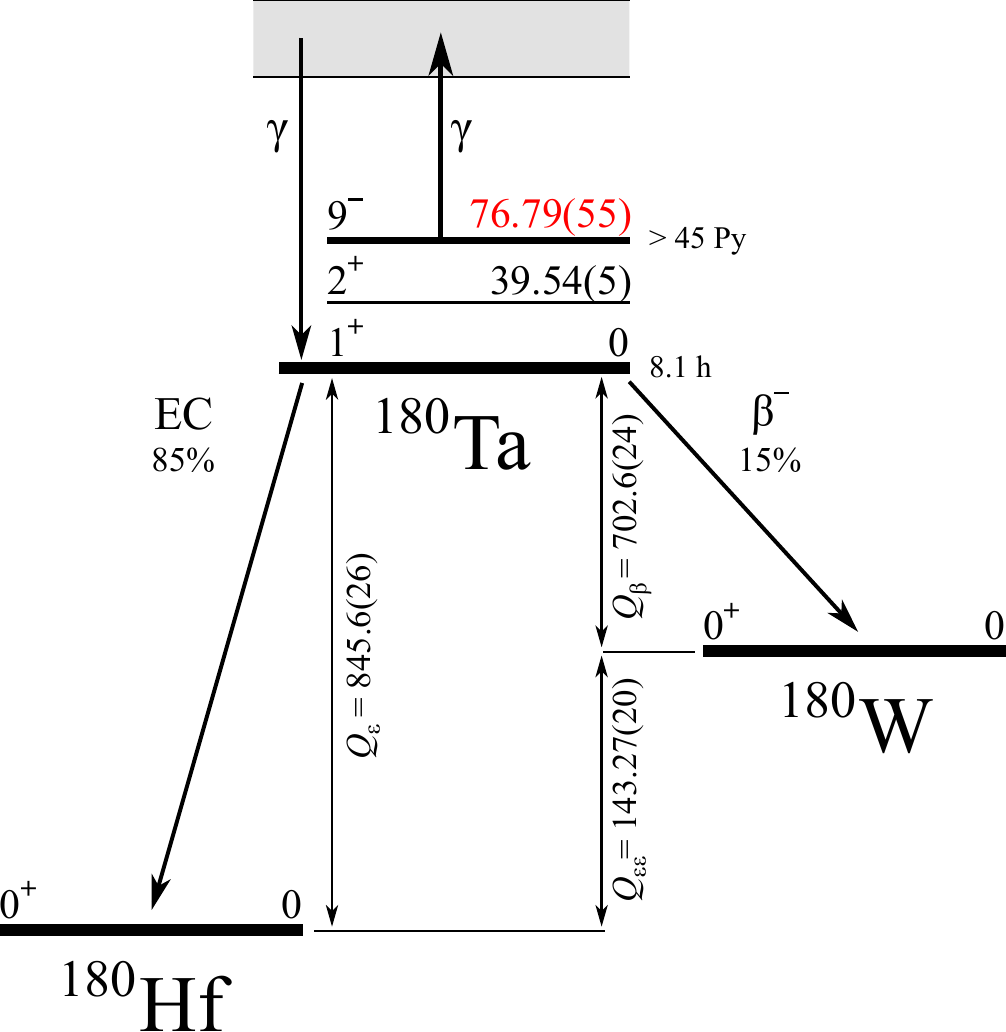}
\caption{(colour online) Decay scheme of $^{180}$Ta. Energy levels in $^{180}$Ta and $Q$ values \cite{AME-20} are given in keV. Due to the high spin difference, a depopulation of the naturally abundant 76.79 keV state in $^{180m}$Ta can only occur by a photoexcitation into resonance intermediate states which have a decay branch into the ground state. The ground state, in turn, decays either by $\beta^-$ decay to $^{180}$W or by electron capture ($EC$) to $^{180}$Hf.}
\label{fig:scheme}
\end{figure}

Due to the large spin difference between the $1^+$ ground state and the $9^-$ isomer, the internal transition to the ground state is not possible. The ground state decays via $\beta^-$ decay to $^{180}$W and via electron capture ($EC$) decay to $^{180}$Hf (see Fig.~\ref{fig:scheme}). For the $^{180m}$Ta isomer, the beta decay branches are non-existent or extremely small as there are no $8^-$, $9^-$, $10^-$  states  within the $Q_{\beta}-/Q_{EC}$ window.

$^{180m}$Ta has been studied in detail from many perspectives. The synthesis of $^{180m}$Ta, the rarest stable isotope in nature, is still not well understood. 
Thermal excitation and de-excitations are also affected by nuclear structure of $^{180m}$Ta, which is a prolate nucleus with the projection of the spin on the nuclear deformation axis $K=9$. The bands built on the $K^{\pi} = 9^-$ isomer and $K^{\pi} = 1^+$ ground state have been studied in detail \cite{Walker,Dracoulis1998,Saitoh1999,Dracoulis2000}.
At high stellar temperatures, the isomer can be depopulated via thermal excitation to higher-lying states, which eventually de-excite to the short-lived ground state. The production of $^{180m}$Ta via the slow neutron-capture process ($s$-process) as well as $p$- and $\nu$-processes has been studied \cite{Kappeler2011,Mohr}. 
It has been proposed as a good candidate for the development of a gamma laser \cite{Walker}. $^{180m}$Ta has also potential to be used in experiments devoted to the search for dark matter \cite{Pospelov}.

So far, a direct decay of $^{180m}$Ta has never been observed. Its excitation energy has been evaluated based on various nuclear reactions, resulting in values 75.3(14) keV \cite{AME-20} and 77.2(12) keV \cite{NDS-15}. There are several examples in the literature, where an evaluated value has turned out to be in a strong disagreement with the result achieved by direct Penning-trap mass measurements (for example, a $10\sigma$ discrepancy was found for $^{102}$Pd--$^{102}$Ru in \cite{Nesterenko-2019}). Thus, a direct mass measurement is of outermost importance. 
In this work, we have used Penning-trap mass spectrometry to reliably measure the key spectroscopic parameter of the isomer -- its excitation energy -- as a mass difference between the isomeric and ground state of $^{180}$Ta.

\section{Experimental method and results}

The measurement was performed with singly charged $^{180}$Ta$^+$ and $^{180m}$Ta$^+$ ions with the Penning-trap mass spectrometer JYFLTRAP \cite{Eronen2012} at the Ion Guide Isotope Separator On-Line (IGISOL) facility \cite{Moore2013}. Both $^{180}$Ta$^+$ and $^{180m}$Ta$^+$ ions were simultaneously produced using nuclear reactions with a 40 MeV proton beam impinging into a $^\text{nat}$Ta target with a thickness of about 5~mg/cm$^2$. The experimental reaction cross section for a production of $^{180}$Ta at this proton energy is a few hundred mb \cite{Uddin}, and a similar cross section is expected for the production of $^{180m}$Ta according to TALYS simulations \cite{talys}. The reaction products were thermalized in the IGISOL gas cell filled with helium gas at a pressure of about 150 mbar. A large fraction of ions ends up as singly-charged. The ions are extracted from the IGISOL gas cell and guided via a sextupole ion guide \cite{Karvonen2008} to a high-vacuum region. Then they were accelerated to 30 keV energy with subsequent mass separation by a $55^{\circ}$ dipole magnet. The continuous ion beam with the selected mass number of $A=180$ was injected into a gas-filled radio-frequency quadrupole \cite{Nieminen2001}, where it is converted into compact ion bunches. Finally, the ion bunches are guided to the JYFLTRAP double Penning trap mass spectrometer.

\begin{figure}[htb]
\centering
\includegraphics[width=0.49\textwidth]{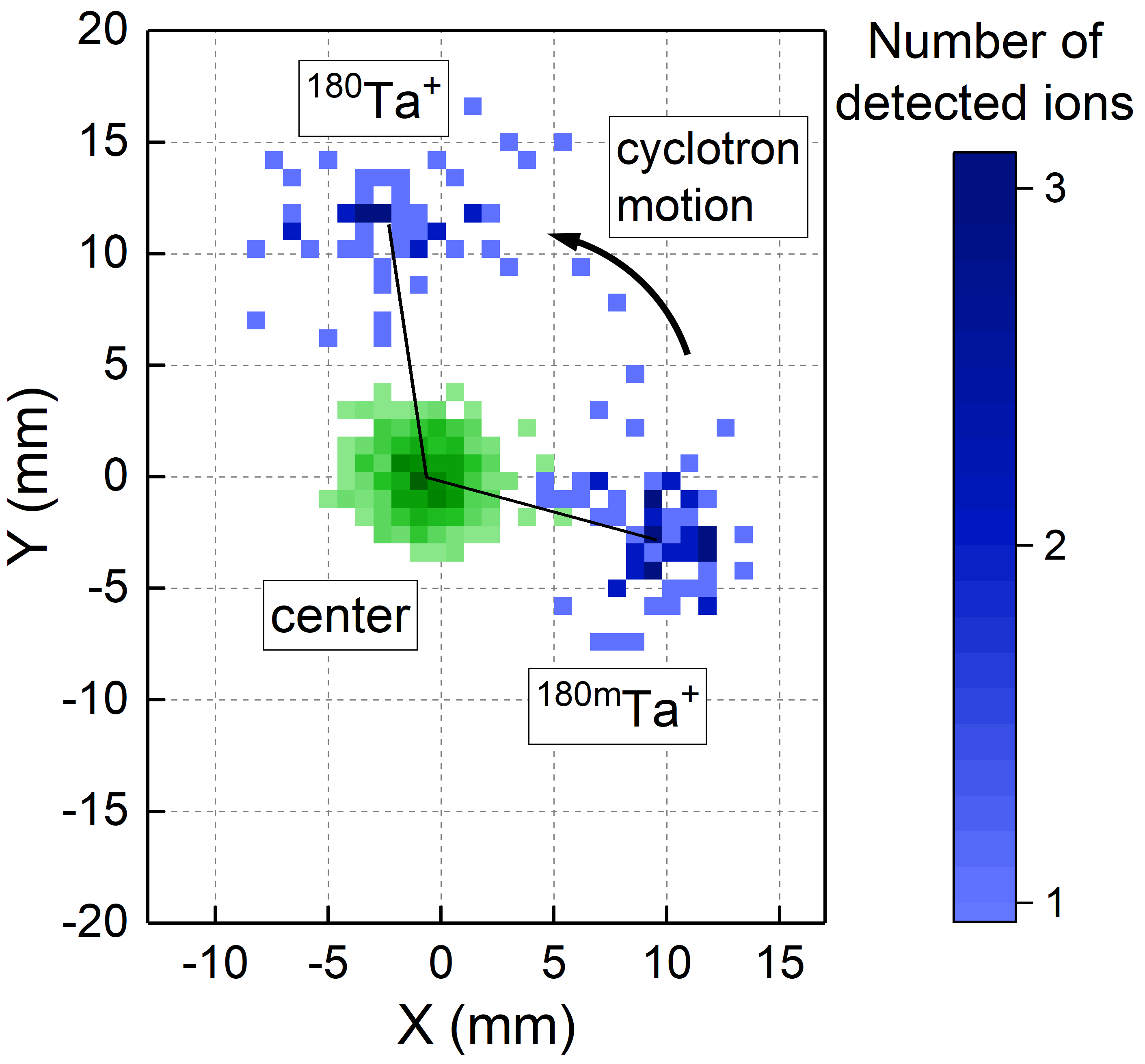}
\caption{(colour online) Projection of the cyclotron phase image (blue bins) and the trap center (green bins) onto the position-sensitive MCP detector for the $^{180}\text{Ta}^+$ and $^{180m}\text{Ta}^+$ ions in a single cyclotron frequency measurement with the PI-ICR method. The phase accumulation time $t_{acc}$ was about of 1125 ms.}
\label{fig:piicr}
\end{figure}

In the mass spectrometer the ions were cooled, centered and purified in the first preparation Penning trap by means of the mass-selective buffer gas cooling technique \cite{Savard1991}. The process requires approximately 300 ms. Since the mass difference between the isobaric contaminant $^{180}\text{W}^+$ and the ion of interest $^{180}\text{Ta}^+$ is only about 700~keV/c$^2$, corresponding to about 2.5 Hz in cyclotron frequency difference, the buffer gas cooling technique is unable to resolve them. To get rid of $^{180}\text{W}^+$ ions, after cooling in the preparation trap the ions were transported to the second (measurement) trap where a 200-ms dipolar pulse at the corresponding reduced cyclotron frequency $\nu_+$  was applied to excite these ions to large cyclotron radii, larger then the radius of the diaphragm between the traps. Then the ions were sent back to the first trap, and only ions of interest that remain cooled at the trap center passed the diaphragm and were subject to the final cooling that takes 160 ms. Finally, only $^{180}\text{Ta}^+$ ions, both the ground and the isomeric state, were transported to the measurement trap for the final determination of their mass. Tantalum is a highly reactive chemical element and easily forms $^{180}\mathrm{T}^{16}\mathrm{O}^+$ compound in the measurement trap by interactions of the tantalum ions with  residual gas particles. To get rid of this constantly forming tantalum oxide contamination, a continuous dipolar excitation at the corresponding $\nu_+$ frequency was constantly applied over the entire duration of the measurement. Since the masses and thus the frequencies of $^{180}\mathrm{T}^{16}\mathrm{O}^+$ and $^{180}\mathrm{Ta}^+$ ions differ significantly, the continuous cleaning excitation pulse does not affect the actual mass measurement of $^{180}\text{Ta}^+$ states.

In Penning trap mass spectrometry, the mass $m$ of an ion is determined by measuring the ion's cyclotron frequency $\nu_c$ 
\begin{equation} \label{eq:qbm}
\nu_{c} = \frac{1}{2\pi}\frac{q}{m}B,
\end{equation}
where $q/m$ is the ion's charge-to-mass ratio and $B$ is the magnetic field strength. The cyclotron frequency is measured with the phase-imaging ion-cyclotron-resonance (PI-ICR) technique that is  described in detail in \cite{Eliseev2013, Eliseev2014, Nesterenko2018}. Here only the main concept and the details specific to the presented measurement are given. The PI-ICR method is based on the observation of a phase evolution of the magnetron and cyclotron ion motions with frequencies $\nu_-$ and $\nu_+$, respectively,  by projecting the ion's position in the trap onto a position-sensitive detector. The projection of the position of the ion that performs pure magnetron and cyclotron motions are called $\nu_-$ and $\nu_+$ phase images (spots), respectively. The positions of the phase images are described with polar angles $\alpha_-$ and $\alpha_+$, respectively, with respect to the trap center. The cyclotron frequency $\nu_c$ calculated as a sum of two radial frequencies \cite{Gabrielse-2010} is determined from the angle $\alpha_c = \alpha_+ - \alpha_-$ between these two phase images as
\begin{equation} \label{eq:alpha}
\nu_{c}= \nu_{-} + \nu_{+} = \frac{\alpha_c + 2 \pi n} {2 \pi t_\mathrm{acc}},
\end{equation}
where $n$ is the full number of revolutions during the phase accumulation time $t_\text{acc}$. For $\nu_c$ frequency measurement of a certain ion species the phase spots and the center spot were alternately accumulated with a period of about 3 minutes. In turn, measurements between the two ion species were alternated after 4 such rounds, i.e with a period of about 12 minutes.

\begin{figure}[htb]
\centering
\includegraphics[width=0.5\textwidth]{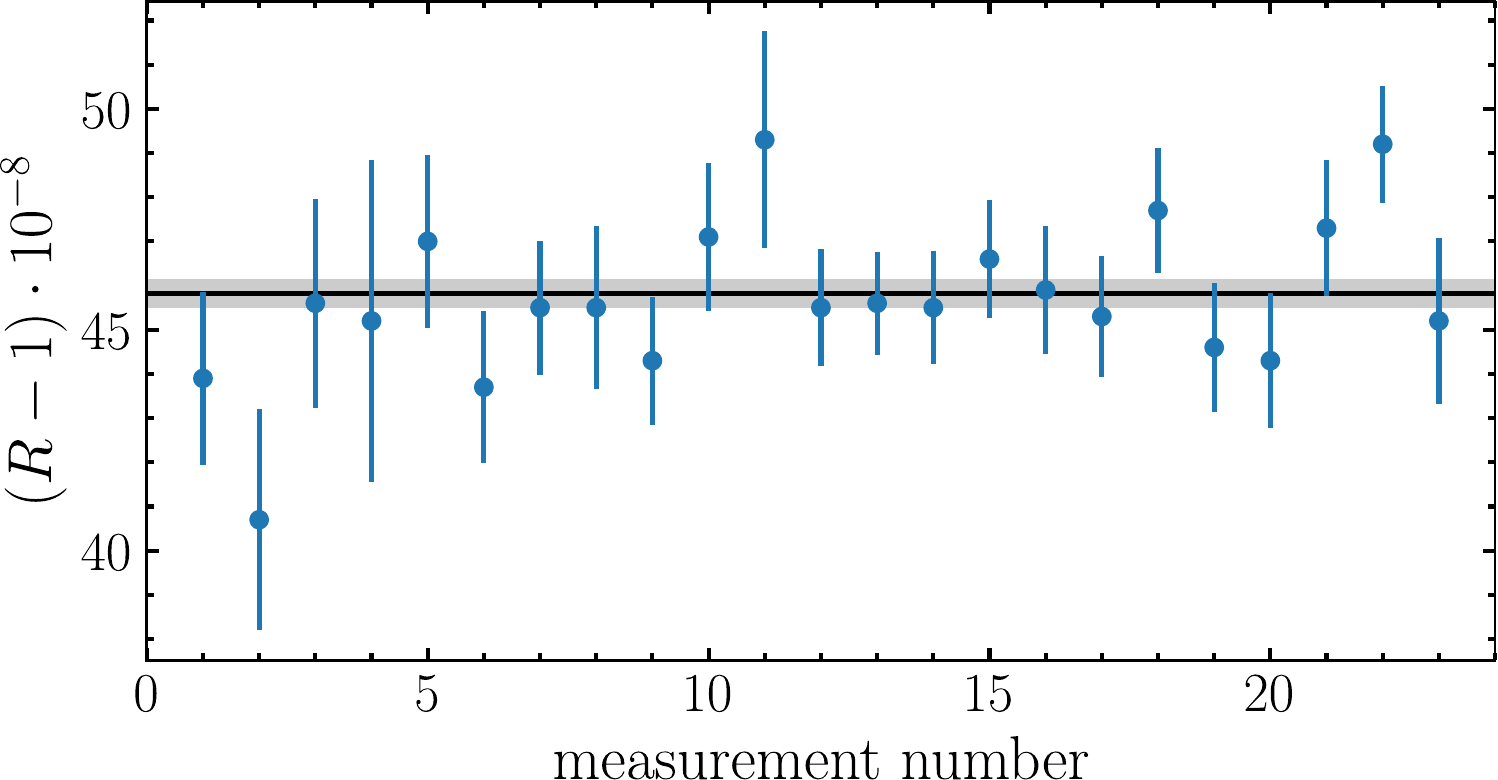}
\caption{(colour online) Cyclotron frequency ratios $R = \nu_c(^{180}$Ta$^+) / \nu_c(^{180m}$Ta$^+)$ determined for each 2-hour measurement period. The grey band represents the total 1$\sigma$ uncertainty of the weighted mean frequency ratio $\bar{R} = 1.0000004581(33)$.}
\label{fig:ratios}
\end{figure}

The phase accumulation time $t_\text{acc}$ of 950~ms has been used to resolve the ground and isomeric state in $^{180}$Ta. As a crosscheck, a part of the data was acquired with $t_\text{acc} = 1125$~ms (Fig.~\ref{fig:piicr}) to ensure that the cyclotron spots on the detector are not contaminated by any possible isobaric impurities. Since both states are simultaneously produced, have similar yields, and cannot be efficiently separated prior to the actual PI-ICR measurement, for minimization of the cyclotron frequency shift due to an ion-ion interaction between these different ion species, the measurement was carried out with a low count rate of approximately one detected ion per five bunches. The measurement settings for both the ground and isomeric state are the same except for the delay and frequency of the conversion pulse, which are adjusted to minimize angle $\alpha_c$ between the magnetron and cyclotron phases for each state. The average angle $\alpha_c$ in the measurements did not exceed a few degrees to minimize the systematic shifts due to the distortion of the phase projections.

Due to the intentionally low count rate, a couple of hours were required to collect reasonable statistics for an individual $\nu_c$ determination. Thus, the total measurement time was divided into 23 approximately 2-hour periods. For each period the $\nu_c$ frequency for both ion species was determined and a single cyclotron frequency ratio $R = \nu_c(^{180}\mathrm{Ta}^+) / \nu_c(^{180\text{m}}\mathrm{Ta}^+)$ was calculated, as shown in Fig.~\ref{fig:ratios}. The final cyclotron-frequency ratio $\bar{R} = 1.0000004581(33)$ was obtained as the weighted mean of these single ratios with the uncertainty taken as the larger of the internal and external statistical uncertainties \cite{Birge1932}. The associated Birge ratio was about a unity. The systematic uncertainties specific to the PI-ICR method are discussed in \cite{Nesterenko2021}. The mass-dependent systematic effects for mass doublets are negligible \cite{Roux2013}, thus the statistical uncertainty is dominant in the final uncertainty. 

The excitation energy $E^*$ of $^{180m}$Ta was calculated from the cyclotron frequency ratio as
\begin{multline} \label{eq:Q}
E^* = \left( M(^{180m}\text{Ta}) - M(^{180}\text{Ta}) \right) c^2 = \\ ( \bar{R} - 1 ) \left( M(^{180}\text{Ta}) - m_e \right) c^2,
\end{multline}
where $M$($^{180}$Ta) and $M$($^{180m}$Ta) are the atomic masses of $^{180}$Ta and $^{180m}$Ta, respectively, and $m_e$ is the electron mass. The uncertainty of the tantalum mass value $\delta M(^{180}\text{Ta})=2.1$ keV/$c^2$ \cite{Wang2021} does not affect the precision since the first term $(\bar{R} - 1) < 10^{-5}$. For the same reason, the binding energy of the valence electron can be neglected. Ultimately, the determined excitation energy of $^{180\text{m}}$Ta is $E^* = 76.79(55)$~keV.

\section{Conclusion}

$^{180\text{m}}$Ta is an unusual isomeric state. It is the only excited nuclear state present in nature and the longest-lived isomer. It is also a state of potential interest in various fields of physics. However, its main characteristics, namely the excitation energy, had not been measured directly. In this work, we have measured the cyclotron frequency ratio between the $^{180}\text{Ta}^+$ ground-state and isomeric-state ions in a single experiment at the Penning-trap facility JYFLTRAP. 
With the obtained cyclotron frequency ratio, $R = 1.0000004581(33)$, we calculated the excitation energy of the longest-lived isomer $^{180m}\text{Ta}$ to be $E^* = 76.79(55)$~keV. Our new value is a factor of two more precise and in a good agreement with the previously known literature values \cite{AME-20,NDS-15} evaluated on the basis of various nuclear reactions with $^{180}\text{Ta}$. 
The new accurate value improves the precision of the excitation energy and related reaction $Q$ values relevant for stellar nucleosynthesis \cite{Mohr, Walker}, driven gamma emission in $^{180}$Ta and gamma laser development \cite{Carroll-01}, and to a dark matter search \cite{Lehnert}.

\section*{Acknowledgements}
The experiment was performed in 2021. This work has been supported by the European Union’s Horizon 2020 research and innovation programme under grant agreements No 771036 (ERC CoG MAIDEN) and No 861198–LISA–H2020-MSCA-ITN-2019 and by the Academy of Finland under projects No 295207 and 327629.

\bibliography{mybibfile}

\end{document}